\documentclass[%
 reprint,
 amsmath,amssymb,
 aps,
]{revtex4-2}

\usepackage{graphicx}
\usepackage{dcolumn}
\usepackage{bm}

\makeatletter
\def\fourvdots{\vbox{\baselineskip1\p@ \lineskiplimit\z@\kern6\p@\hbox{.}\hbox{.}\hbox{.}\hbox{.}}}
\makeatother
\usepackage{hyperref}

\usepackage{physics}
\usepackage{bbm}
\usepackage[dvipsnames]{xcolor}

\begin{document}

\preprint{APS/123-QED}

\title{Understanding and embracing imperfection in physical learning networks}

\makeatletter
\renewcommand*{\@fnsymbol}[1]{\ifcase#1\or $\dagger$\or $\ddagger$\or iii\or iv\or v\or vi\or vii\or viii\or ix\or x\else\@ctrerr\fi}
\makeatother

\author{Sam Dillavou*$^{1}$}%
\email{dillavou@sas.upenn.edu}
 \thanks{*Equal contribution.}
\author{Marcelo Guzman*$^{1}$}
 \email{mguzmanj@sas.upenn.edu}
 \thanks{*Equal contribution.}
\author{Andrea J. Liu$^{1,2}$}
\author{Douglas J. Durian$^{1}$}
\affiliation{$^1$Department of Physics and Astronomy, University of Pennsylvania, Philadelphia, PA 19104, USA.
}%
\affiliation{$^{2}$Santa Fe Institute, NM 87501, USA.}

\date{\today}

\begin{abstract}
Performing machine learning with analog signals offers advantages in speed and energy efficiency, but sensitivity to component and measurement imperfections often foils training without a system-specific companion digital model.
Here we take a different perspective, accepting and characterizing these inherent imperfections and ultimately overcoming them without digital models.
We train an analog network of self-adjusting resistors---a contrastive local learning network---for multiple tasks, and observe limit cycles and scaling behaviors that limit precision, erase memory of previous tasks, and are absent in `perfect' systems.
We develop an analytical model capturing these phenomena as a consequence of an uncontrolled learning bias continuously modifying the underlying representation of learned tasks, reminiscent of \textit{representational drift} in the brain. Finally, we introduce and demonstrate a system-agnostic training method that greatly suppresses these effects.
Our work points to a new, scalable analog approach that eschews precise modeling and instead thrives in the mess of real systems.
\end{abstract}

\maketitle

\section{Introduction}
Digital neural networks are expressive and malleable mathematical functions comprised of many simple and tunable nonlinear functions. Training these networks is typically accomplished using iterative gradient-based methods, which implicitly rely on extremely precise \textit{modeling} of the network and \textit{adjustment} of its tunable parameters.
In digital networks, achieving such precision is straightforward because the model \textit{is} the network, and simulated adjustments can be calculated and performed using any number of bits. However, when this precision is intentionally removed, for example by biasing the updates, training typically fails to achieve an acceptable accuracy~\cite{ghosh_how_2023_2}.
This highlights a problem with applying these training methods to physically-instantiated analog networks, where modeling, measurement, and operation of components will invariably be imperfect at some level.

This simulation-reality gap is a major obstacle for many forms of \textit{neuromorphic computing}~\cite{momeni_training_2024, haensch2023compute} and \textit{physical learning}~\cite{stern_learning_2023}, fields aimed at leveraging physical processes or entire physical systems to perform machine learning tasks. These approaches offer enormous potential advantages in energy-efficiency, speed, and other factors like physical robustness~\cite{dillavou_demonstration_2022}, but cannot assume pristine digital conditions. As a result, experiments modeled after digital ML systems typically use simulation and/or incorporate digital processing~\cite{wright_deep_2022,onodera_scaling_2024, pai_experimentally_2023, xue_fully_2024, wang_reinforcement_2019}, limiting their ability to scale up. Alternatively, frameworks have been proposed for training physical systems that minimize energy or power~\cite{movellan_contrastive_1991,scellier_equilibrium_2017,stern_supervised_2021,song_inferring_2024,li_training_2024, martin_eqspike_2021, pashine_local_2021}. Here, physical perturbation endows real signals with gradient information, avoiding differentiation of a model. However, imperfect measurements and adjustments of network parameters have prevented experimental demonstrations~\cite{dillavou_demonstration_2022,wycoff_desynchronous_2022, stern_continual_2020,pashine_local_2021} from reaching the accuracy and expressivity of their simulated counterparts~\cite{stern_supervised_2021, martin_eqspike_2021,scellier_energybased_2023a}, even when digital training is incorporated~\cite{li_training_2024,martin_eqspike_2021}.

Biology has proven that learning complex tasks at scale is possible. For example, brains learn and function despite imperfect components, environmental variations, and even substantial damage.  Understanding how the brain overcomes these hurdles is extremely difficult, though obviously it does not succeed by using a companion digital simulation of itself to calculate its adjustments. To understand and ultimately overcome the impact of imperfection on the training of physical networks, it is fruitful to focus on simple systems that learn without digital aid. 

Here we observe, characterize, and ultimately suppress the effects of component imperfection and variability in a physically-constructed self-learning system. Specifically, we study a \textit{Contrastive Local Learning Network}~\cite{dillavou_machine_2024} (CLLN), an analog network in which each edge is a variable nonlinear resistor that autonomously self-adjusts according to a local rule. Such systems learn precise global functionality without digital or central computation. In training our CLLN alternatively between two different tasks, we observe non-equilibrium steady-state dynamics instead of convergence to a fixed point as seen in simulations of CLLNs. 
We introduce a model amenable to theoretical analysis and show that these dynamics appear when \textit{any} bias is present in the adjustment of parameters.
Importantly, biases in measurement, calculation, and application are generic and unavoidable features of analog physical hardware~\cite{haensch2023compute}.
Finally, we use these insights to develop a modified training protocol, \textit{``overclamping"}, which suppresses the effects of bias (and noise, though negligible in our experiments) and drastically improves experimental results on classification tasks. 
Because this technique embraces the inherent imperfection of analog systems---requiring no digital modeling, additional measurements, or hardware improvements---it is not system-specific and can be applied broadly to improve training of physical systems.

\begin{figure*}[t]
\includegraphics[width=17.4cm]{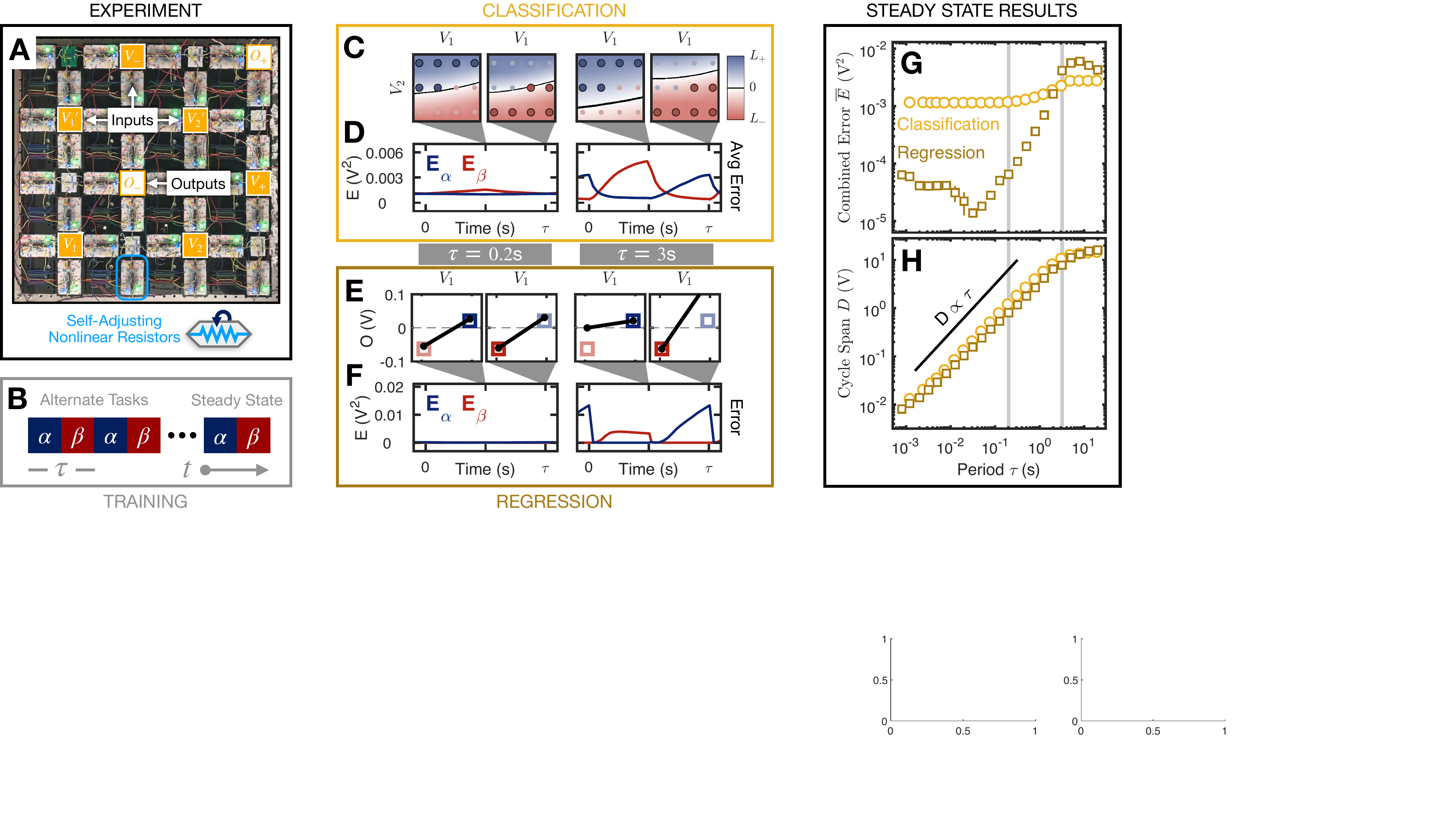}
\caption{\label{fig:fig1} \textbf{Training for Two Tasks} \textbf{A} Image of the Contrastive Local Learning Network (CLLN) decorated with chosen input (yellow) and output (white) node locations. The network is a grid of self-adjusting nonlinear resistors -- specifically n-channel enhancement MOSFET transistors -- and nodes are simply connections between edges. Two copies of the network are constructed (they coexist on the same breadboards) to run the Free and Clamped states (see text). The system output is the difference $O = O_+-O_-$ between two chosen nodes, and variable inputs $V_1$ and $V_2$, their inverses $V_i' = V_+-V_i$, and two constant values $V_+ = 0.435$~V and $V_- = 0.018$~V are imposed as voltage boundary conditions. The differential output and inverted inputs both give the system access to negative relationships between input and output, permitting more complex functionality.  \textbf{B} Training protocol. Two tasks $\alpha$ (blue) and $\beta$ (red) are alternatively trained for time $\tau/2$ until reaching a steady state. \textbf{C} Binary classification. We treat each class as a separate task. Background color is system output, labels $L_\pm =\pm$63~mV, and circles are training data. Steady state results are shown for $t=\tau/2$ and $t=\tau$ for $\tau=$ 0.2~s and 3~s. 
\textbf{D} Average error for each task (class) $E_\alpha$, $E_\beta$ over time in steady state.
\textbf{E} Regression with two data points, separated into two tasks. $V_2$ is held at $V_+/2$ for this task. Labels are squares, output is black line, and panels are paired as in C. 
\textbf{F} Limit cycle error as in D. 
\textbf{G} Combined steady state error $\overline E$ vs $\tau$ for classification (circles) and regression (squares) experiments. Example times from C-F are indicated. \textbf{H} Same as G but for the cycle span $D = \lVert \mathbf{G}(\tau/2) - \mathbf{G}(\tau) \rVert$, that is, the straight-line distance the system travels in a half-cycle.
}
\end{figure*}

\section{Results}

\subsection{Description of the experimental setup}
We analyze the dynamics of a Contrastive Local Learning Network (CLLN), 
an experimental system described extensively in~\cite{dillavou_machine_2024} and briefly here. The system is an analog electronic network of self-adjusting nonlinear resistors, where inputs are represented by enforced voltage boundary conditions on select nodes, and physics performs the `computation' of outputs by enforcing Kirchhoff's current and voltage laws.
The CLLN implements the Coupled Learning (CL) rule~\cite{stern_supervised_2021} that specifies the adjustment of each nonlinear resistor based only on local voltage drops. This rule, closely related to Equilibrium Propagation~\cite{scellier_equilibrium_2017} and Contrastive Hebbian Learning~\cite{movellan_contrastive_1991}, has been used successfully to train physical systems to perform nontrivial tasks such as nonlinear classification and regression~\cite{dillavou_machine_2024, dillavou_nonlinear_2023}. 

The implementation of CL for analog electronic networks is as follows. First, a set of nodes are selected as inputs and output(s). A desired input-output response (task) is then prescribed by selecting training data points $[\vec V_i, L_i]$ where $\vec V_i=V_{1,i}, V_{2,i},...$ are the voltages imposed at the input nodes and $L_i$ is the desired output node voltage (label). In this work we consider only tasks with a single output voltage, but our framework and findings easily extend to tasks with multiple outputs~\cite{dillavou_machine_2024}. The system is then exposed to two electronic states: the \textit{Free} state where the inputs $\vec V_i$ are imposed and the network produces output $O_i^F$, and the \textit{Clamped} state where the inputs \textit{and} `nudged' output $O_i^C$,
\begin{equation}
    O_i^C = O_i^F + \eta \delta_i
    \label{clamping}
\end{equation}
are imposed, where $\eta \in (0,1]$ is a hyperparameter, and $\delta_i = L_i-O_i^F$ is the signed error for datapoint $i$.

The elements of the network then adjust their conductances based on the voltage drops they experienced in these two states, $V_F$ and $V_C$.  In our CLLN, shown in Fig.~\ref{fig:fig1}A, these elements are MOSFET transistors in the Ohmic regime whose conductance $K$ is controlled by a gate voltage $G$ held on a capacitor, specifically
\begin{equation}
    K = S(G - V_T - \overline V)
    \label{conductance}
\end{equation}
where $S \approx 8 \times 10^{-4} \ (\textrm{V} \text{ ohms})^{-1}$, $V_T \approx 0.7$~V, and $\overline V$ is the average voltage of the two nodes in contact with the edge. 
Despite being in the Ohmic regime, Eq.~\eqref{conductance} leads to a non-linear relation between the current across the transistor and its terminal voltages, as the conductance itself depends on the average voltage $\overline V$---this is the source of nonlinearity in our system~\cite{dillavou_machine_2024}.
We also emphasize that while our experimental network has a grid structure, it can accommodate any architecture.
Moreover, unlike crossbar-based analog implementations~\cite{haensch2023compute}, our CLLN does not rely on accurately implementing matrix-vector-multiplication needed for feed-forward networks; it is a fully recurrent network implementing local learning rules.


The analog circuitry on each edge then adjusts the gate voltages according to
\begin{equation}
    \Delta G = t_h \times A_0 \Big (V_F^2 - V_C^2\Big ),
    \label{expRule}
\end{equation}
where $t_h$ is the amount of time the system is allowed to charge the capacitor (learning `on') and 
$A_0\approx 2.5\times 10^3 ~(\text{V s})^{-1}$, determined by choice of electronic components, see Materials and Methods. Our experiment uses twin networks to separately but simultaneously impose the \textit{Free} and \textit{Clamped} boundary conditions, and the two networks are updated in tandem so that the gate voltages on the twin edges are always the same.

This update rule and its success in training were derived and demonstrated in \cite{stern_supervised_2021}, and specifically for this system in \cite{dillavou_machine_2024}. Simply put, Eq.~\eqref{expRule} evolves the edge conductances to minimize the difference in power dissipated between the free and clamped states, which we call the \textit{Contrastive Function} $\mathcal C=\mathcal P_C-\mathcal P_F$.
$\mathcal C$ is used in~lieu of a more standard loss function because it can be minimized using local rules. Minimizing $\mathcal C$ brings the free and clamped states into alignment, which also minimizes error $|\delta_i| \rightarrow 0$ and evolution (in theory) stops.
This description holds for tasks with multiple data points, provided they are rapidly cycled as the system learns~\cite{dillavou_machine_2024}.

We explore the effects of physical imperfection by training the system alternatively for two tasks $\alpha$ and $\beta$. We choose a period $\tau$, training the system for task $\alpha$ for time $\tau/2$, then task $\beta$ for time $\tau/2$, and then repeating these two steps over and over until the gate voltages reach a steady state, as shown schematically in Fig.~\ref{fig:fig1}B. 
Note that when a task involves multiple data points, they are cycled in order, each exposed to the network for time $t_h=100~\mu$s until $\tau/2$ has elapsed. 


\subsection{Experimental Results}

Fig.~\ref{fig:fig1}C and D illustrate the alternate training in binary classification, where each class corresponds to a task. For classification, the network is trained using the normal protocol, but clamping is removed ($\eta \rightarrow 0$) when a datapoint is `very correctly classified', here meaning that its output is beyond the label on the correct side of the boundary. This effectively creates a shifted hinge loss, see Materials and Methods for details.
Unsurprisingly, we find that for sufficiently fast switching between the tasks (small periods $\tau$), the system finds a steady state where errors (squared hinge loss) and outputs are relatively constant, as shown for $\tau = 0.2$~s.
However, as $\tau$ grows, a clear cyclic behavior emerges, with error for the task not being trained growing as the other shrinks. Note that the error for the `off' task continues to evolve even when the `on' task error finds a steady value, clearly seen in the $\tau=3s$ example.

We repeat this cyclic protocol for a pair of far simpler (regression) tasks, where each task is a single desired input-output relationship (data point). The same trends emerge, with minimal movement at small $\tau$ but tradeoffs between task errors (mean-squared errors) at larger $\tau$. Here again, the longer the system trains for one task, the more its performance degrades for the other, resembling the phenomenon of catastrophic forgetting in machine learning~\cite{kirkpatrick2017overcoming}---where memories of old tasks are erased by learning new ones---although here it is driven by physical imperfections, rather than training schedule alone.

We find that in steady state, the combined error across both tasks and half-periods, $\overline E$ (equations in Materials and Methods), follows a similar pattern in both the classification and regression tasks, as shown in Fig.~\ref{fig:fig1}G.
Specifically, the error is independent of the training period $\tau$ for both short and long cycles, but in an intermediate regime it increases monotonically with $\tau$.

By tracking the gate voltage values $\mathbf G =(G_1,G_2,...)$, we characterize the distance the system moves in one half period through the cycle span $D$,
\begin{equation}
    D = \lVert \mathbf{G}(\tau/2) - \mathbf{G}(\tau) \rVert,
\end{equation}
finding a striking convergence between our two very different pairs of tasks.
Fig.~\ref{fig:fig1}H shows that for many decades, $D\propto \tau$, including the short-period regime where the error $\overline E$ plateaus. This scaling rolls off and $D$ plateaus at high $\tau$. 

All of the behaviors noted in Fig.~\ref{fig:fig1} were unanticipated. 
For a more precise network (\textit{e.g.} a simulation of the network) learning these simple tasks would not display \textit{any} steady-state cycles, but only zero-error fixed points ($\overline E(\tau) =0$ and $D(\tau) =0$). Notably, the transition out of the low-error plateau occurs as $\tau$ eclipses the characteristic timescale of learning ($\tau_0 \approx 18$ ms, set by chosen resistors and capacitors~\cite{dillavou_machine_2024}) times the number of datapoints. This suggests the system `sees' datapoints collectively at small $\tau$ and individually at large $\tau$. However, in a simulation, this would still only affect the transient behavior prior to finding the zero-error solution.
In the next section, we will show that these dynamics stem from \textit{imperfections} in our learning system. 

\subsubsection*{Small Network}
\begin{figure}
\includegraphics[width=8.4cm]{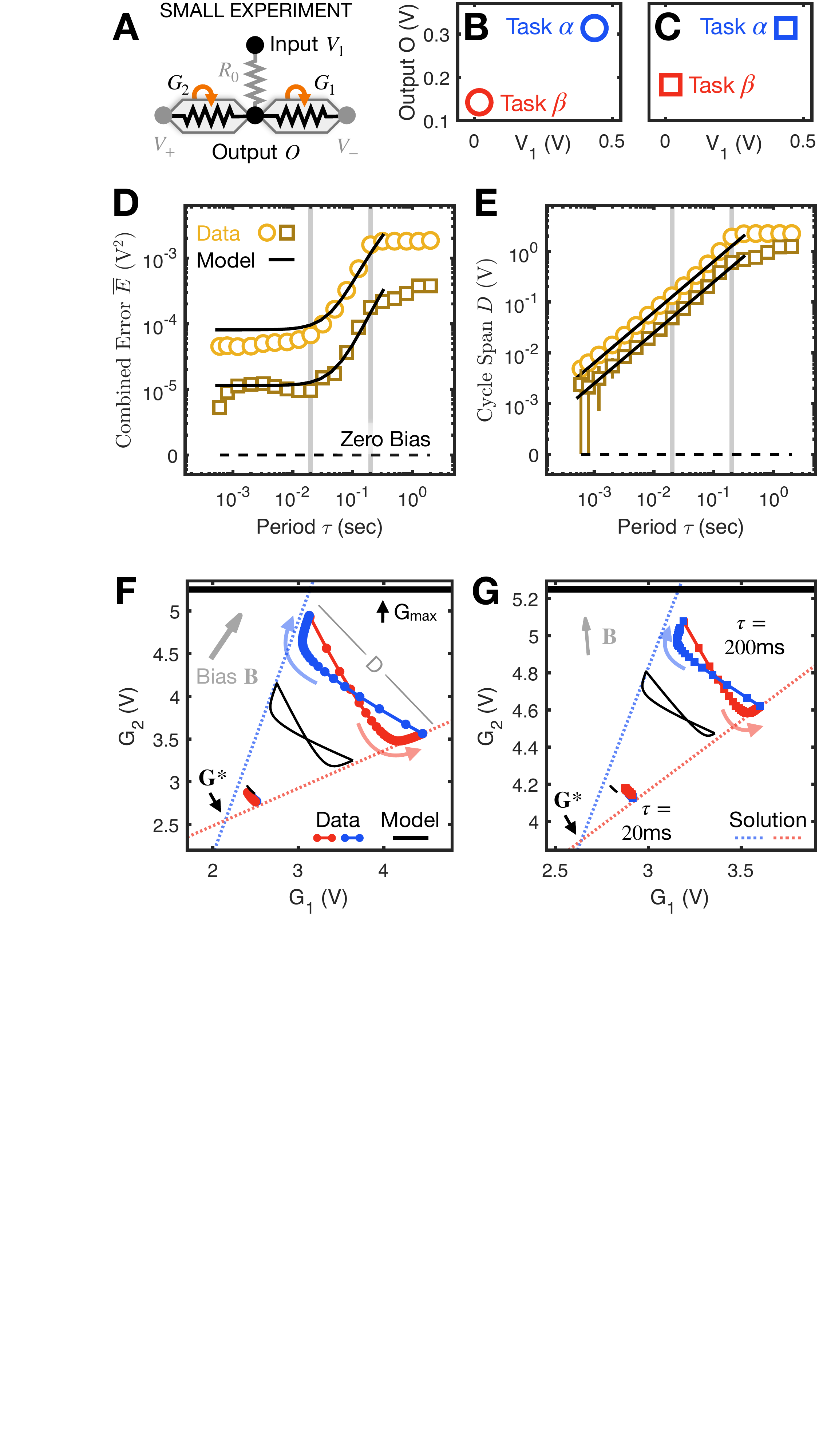}
\caption{\label{fig:fig2} \textbf{Learning Dynamics in a Small Experimental Network} \textbf{A} Schematic of 2+1 edge experiment. \textbf{B, C} Pairs of single data point tasks, differing only by the desired output for task $\beta$. Solid, dashed gray lines connect data points in B,C.
\textbf{D} Combined steady state error $\bar E$ as a function of $\tau$ for task pairs in B (circles) and C (squares). Vertical lines indicate $\tau$ values shown in F and G. Model predictions with no free parameters are shown as solid black lines. 
\textbf{E} Same as D but for Cycle Span $D$. 
\textbf{F, G} Visualization of limit cycles in $G$-space for task pairs in B and C, respectively. Colored points are experimental data for periods $\tau=$ 20 (lower left) and 800~ms (large bow-tie). Faded red and blue arrows indicate cycle direction. Colored dotted lines are individual task solution lines and their intersection is the joint solution $\mathbf{G ^*}$. Gray arrows indicate measured bias $\mathbf{B}$. Thick black lines indicate the parameter maximum $G_{\text{max}}$. Note that edge $G_1$ has been modified to induce a different bias for task C, see Materials and Methods for details.
}
\end{figure}

We now examine a much smaller experimental system that displays similar phenomena, but which can be matched with analytic theory. Specifically, we interrogate a network of two self-adjusting edges and one fixed edge, as shown schematically in Fig.~\ref{fig:fig2}A.
This network has three inputs, two that we hold constant $V_+ =0.435$~V, $V_- = 0.02$~V and one that we treat as our variable input as $V_1=V_+$ or $V_-$. We analyze two scenarios of switching between single-data point tasks (Fig.~\ref{fig:fig2}B and C), wherein task $\alpha$ is the same in both cases. For each single-data point task, there is a line of solutions in $G$ space; these are indicated as the blue and red dotted lines in Fig.~\ref{fig:fig2}F and G for the tasks in B and C, respectively. The intersection point of these solution lines, $\mathbf G^*$, corresponds to the joint solution. 
Following our previous (alternating) training protocol, a `perfect' system will always reach and remain at this point, regardless of the period $\tau$.

In contrast, the 2+1 edge experiment is never observed to settle at this point. Rather, it displays the phenomenology of the full experimental system; the combined error $\overline E(\tau)$ displays two plateaus and a transition (Fig.~\ref{fig:fig2}D), and the cycle span scales linearly $D \propto \tau$ with an upper cutoff (Fig.~\ref{fig:fig2}E). The limit cycles underlying these behaviors are 2-dimensional $\mathbf G =(G_1,G_2)$, as shown in Fig.~\ref{fig:fig2}F and G. 
Instead of settling at the point of intersection of the red and blue dotted lines in Fig.~\ref{fig:fig2}F and G, our system settles into a limit cycle that follows a `bow-tie' trajectory. When training for task $\alpha$, the system rapidly approaches the corresponding solution space then slowly drifts along it. Switching back to task $\beta$, the system exhibits similar behavior. For fast cycles, the span of each bow-tie scales with $\tau$ and its location converges near (but not on) the joint solution, generating a low-error plateau. As $\tau$ increases, the trajectories grow and drift away from this point, eventually hitting the limits of the experimental parameters $G_{\text{max}}$, after which no change is registered by increasing $\tau$, resulting in the high-$\tau$ plateaus in both error and span.

We note that the same trajectories are found regardless of initial conditions, even when the system is initialized at or passes through the joint solution. This repeatability and the smoothness of the dynamics show that they do not stem from noise. Rather, these features, along with the linear cycle span and drift along the solution lines, point to a systematic learning bias.
\subsection{Modeling Imperfection}
To test this hypothesis, we model the experimental learning dynamics by adding a bias term to the original learning rule in Eq.~\eqref{expRule}, $\mathbf B = (B_1, B_2, \cdots)$, whose components represent the scalar biases for each self-adjusting edge, constant in time but generically different across edges. 
In the continuum (small step) limit, the learning dynamics become
\begin{equation}
    \frac{d \mathbf G}{dt} = -\gamma \frac{\partial \mathcal C}{\partial \mathbf G}+\mathbf B,
    \label{eq:learningwithbias}
\end{equation}
where $\gamma$ is an effective learning rate defined by the previously described experimental parameters of Eqs.~\eqref{conductance} and~\eqref{expRule}; $\gamma = A_0/S$.

As seen in the bow-tie trajectories in Figs.~\ref{fig:fig2}F and G, and by examining Eq.~\ref{eq:learningwithbias}, the effects of imperfection are most influential when $\mathbf G$ is close to a minimum of the contrastive function.
In this regime, the learning dynamics become linear and analytically tractable,
\begin{align}
     \frac{d \mathbf G}{dt} &= -\mathbf M(\mathbf G-\mathbf G^*)+\mathbf B,
     \label{eq:learningwithbiaslinear}
\end{align}
where $\mathbf G^*$ is one of potentially many minima of $\mathcal C$ (a learning solution) and $\mathbf M$ is proportional to its Hessian evaluated at that point, with components $M_{ij}=\gamma\partial ^2\mathcal C/\partial G_i\partial G_j\Big|_{\mathbf G^*}$.
Here we discuss the global properties of Eq.~\eqref{eq:learningwithbiaslinear} and refer the reader to the Materials and Methods for details on the derivation and solutions.

First, we note that the eigenvalues of $\mathbf M$ represent the curvatures of the contrastive landscape, and must therefore be non-negative around a minimum ($\mathbf{ G^*}$). Along positive eigenvalue directions, the (overdamped) learning dynamics exhibit exponential decays to learning solutions, slightly shifted due to the presence of bias, eventually reaching a stationary point. In contrast, the learning dynamics along zero eigenvalue (curvature) directions are solely driven by bias, and therefore drift until they hit an experimental limit.

For periodic two-task training, the Hessian and solution spaces switch each half-period. 
At each change, the gate voltages are rapidly confined along the (newly imposed) positive curvature directions, then drift within the nullspace of the new $\mathbf M$, leading to periodic orbits.
Similar non-equilibrum steady state dynamics can also be present without bias, provided the system cannot access a joint solution, \textit{i.e.}~there is no solution point to task $\alpha$ that is also a solution for task $\beta$.  Here, the dynamics will drive the system back and forth between two distinct solutions $\mathbf G_\alpha^*$ and $\mathbf G_\beta^*$.

\subsection{Matching Experiment and Theory}
To test our model we compare its predictions to experimental results. To begin, we first measure the experimental bias $\mathbf B$ using the drift along solution lines and the learning rate $\gamma=A_0/S$ from single-edge tests, see Materials and Methods.
For the linearized dynamics we choose solution points that are close to the observed experimental trajectories. For compatible tasks, the solution points are taken as the common solution: $\mathbf G^*_\alpha=\mathbf G^*_\beta=\mathbf G^*$.
These experimental measurements determine all of the parameters in the model, which displays excellent quantitative agreement with experiment in the combined error $\overline E$ and cycle span $D$. The agreement between theory and experiment is less strong for limit cycles, as shown by the black lines in panels~\ref{fig:fig2}D to G. Limit cycle predictions (Fig.~\ref{fig:fig2}F,G, black lines) are especially sensitive to deviations from the model assumptions, such as the assumption of linear behavior and state-independent bias. Nonetheless, the model displays the same behavior and trends as the experiment (colored cycles).

\begin{figure}
\includegraphics[width=8.4cm]{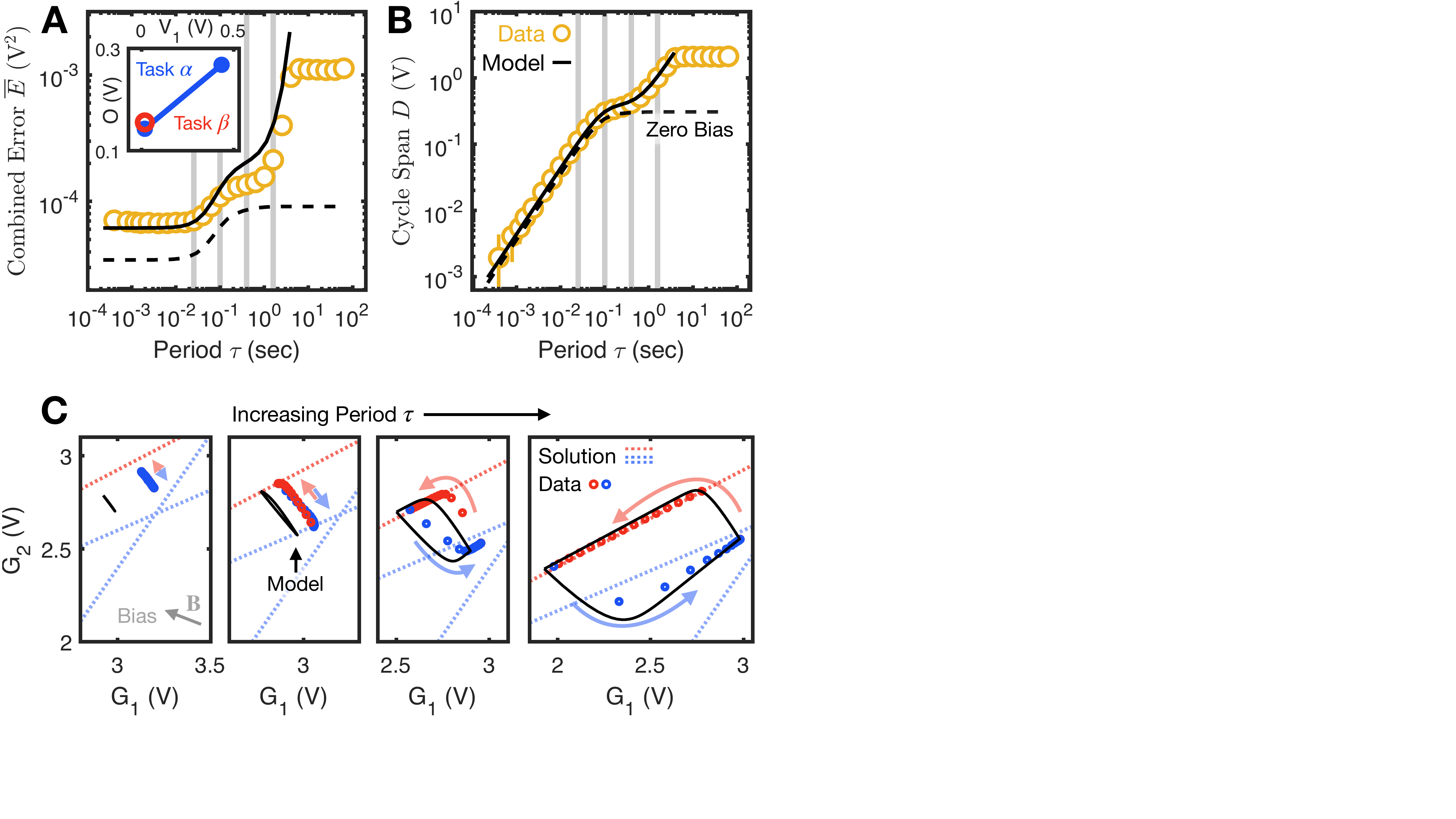}
\caption{\label{fig:fig3} 
\textbf{Incompatible Tasks a Small Experimental Network} \textbf{A inset} A single data point ($\beta$, red) and a two-data point ($\alpha$, blue) task. \textbf{A}  Combined steady state error $\overline E$ as a function of $\tau$. Model prediction with no free parameters is shown as a solid black line. Dashed black line indicates model prediction for zero bias (`perfect') evolution. Vertical lines highlight $\tau$ values shown in C. 
\textbf{B} Cycle span $D$ as a function of $\tau$ with model predictions as in A.
\textbf{C} Visualization of four limit cycles in $G$-space for ($\tau=$ 25~ms, 100~ms, 400~ms, and 1600~ms). Gray arrow indicates measured bias $\mathbf{B}$. Faded red and blue arrows indicate cycle direction. The red dotted line indicates the solution space of task $\beta$, the blue dotted lines the solution spaces of each data point of task $\alpha$, and their intersection is the solution for the full $\alpha$ task.
The X and Y axes for all panels have equal scale, and all Y axes have the same limits. 
}
\end{figure}

For a more stringent test of the model, we alternate between a pair of \textit{incompatible} tasks, that is, two tasks \textit{without} a joint solution ($\mathbf G^*_{\alpha}\neq \mathbf G^*_\beta$).
As shown in Fig.~\ref{fig:fig3}A inset, task $\alpha$ has two data-points---it is effectively a linear regression problem---and task $\beta$ is a single data point that cannot be satisfied simultaneously with task $\alpha$. Each data point for task $\alpha$ has its own solution space (dotted blue lines in Fig.~\ref{fig:fig3}C), as does task $\beta$  (dotted red line). As previously mentioned, the incompatibility between $\alpha$ and $\beta$ leads to periodic trajectories even in the absence of bias, generating a minimum combined error $\overline E$, and a linear scaling in cycle span $D$, as shown by the dashed lines in Fig.~\ref{fig:fig3}A,B.

The dynamics become even richer with the addition of bias. The combined error $\overline E$ exhibits three plateaus, while the cycle span $D$ exhibits a crossover from one regime of linear scaling to another and a high-$\tau$ plateau, as shown in Fig.~\ref{fig:fig3}A and B, respectively. The dynamics underlying these regimes are shown in Fig.~\ref{fig:fig3}C for experiment (colored dots) and the model with no fit parameters (black lines). At small $\tau$ in the absence of bias, the limit cycles would run back and forth between the $\alpha$ solution point $\mathbf{G_\alpha^*}$  (intersection of blue dotted lines) and the nearest point of the $\beta$ solution line, $\mathbf{G_\beta^*}$  (red). Here the only effect of bias is a slight overall shift.
As $\tau$ grows, the limit cycle gets closer to the solutions spaces of each task. As the dynamics slow near the $\beta$ solution, the bias dominates, and drives the gate voltages along the solution line. Upon switching to training for task $\alpha$, the system snaps to the center line between the two $\alpha$ solution lines (one for each datapoint), followed by a slow convergence to (near) $\mathbf{G_\alpha^*}$. Finally, at very large $\tau$ the $G$ values drift until they hit their experimental boundaries, where we truncate our model predictions (end of black lines in Fig.~\ref{fig:fig3}A and B.)

The agreement between theory and experiment suggests that learning bias underlies these rich dynamics. We note, however, that these dynamics are not desired---bias raises the error as the dynamics deviate from a fixed point (for compatible tasks) or a more favorable cycle (for incompatible tasks). 

In terms of the experimental variables, the biased update rule Eq.~\ref{eq:learningwithbiaslinear} can be written for a single edge as
\begin{equation}
    \Delta G= t_h \times \left [ A_0 \Big (V_F^2 - V_C^2\Big ) + B \right ].
    \label{expRule2}
\end{equation}
Imperfections in the physical components can affect the bias $B$ in several ways. Bias can arise from enforcing matching between non-identical twin-networks, inaccurately measuring the free and clamped voltage states, and imprecisely implementing the learning rule. As detailed in Materials and Methods, the latter effect dominates, creating estimated biases of order 2~V/s per edge, in agreement with the measured $B$ values in the previous section. 
Device-to-device variations that cause this effect make \textit{a priori} prediction of the direction of the bias $\mathbf B$ infeasible, so the learning rule per edge cannot be directly corrected.
However, next, we introduce a modification to the training protocol that drastically reduces the effects of bias.

\subsection{Taming Imperfection}

\begin{figure}
\includegraphics[width=8.4cm]{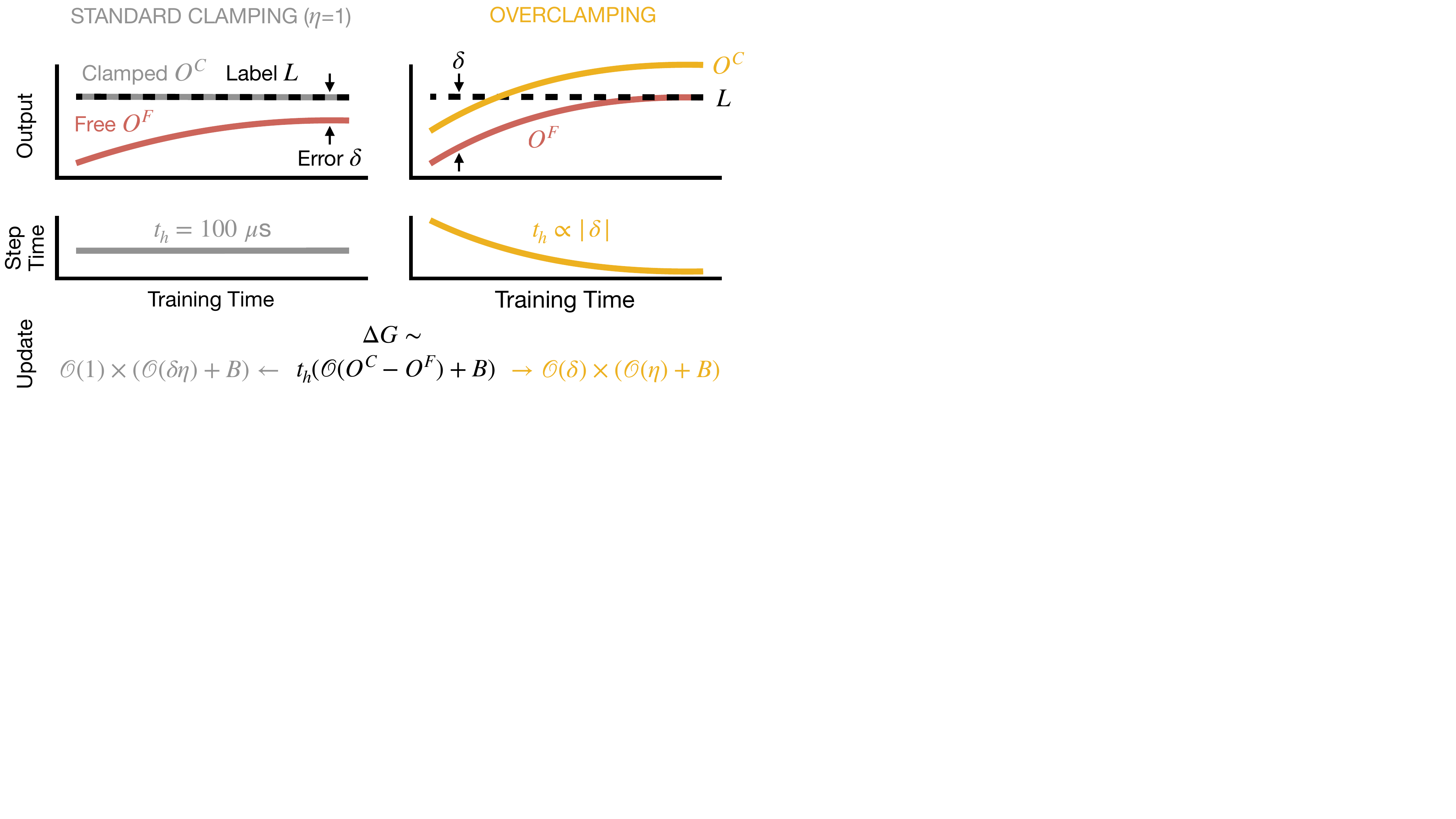}
\caption{\label{fig:fig4} \textbf{Standard vs Overclamping} Schematic example of standard (left, gray) vs overclamping (right, yellow) for a single datapoint. Standard clamping uses $\eta=1$, which results in clamped output $O^C = L$ (label). This technique (with any $\eta$) does not result in the free output settling at the label ($O^F \rightarrow L$) because the signal ($O^C-O^F$) decays with the error $\delta = L-O^F$, and eventually bias dominates. In contrast, overclamping sets the key difference $O^C-O^F$ to an approximate constant, and reduces train step time $t_h$ proportional to the error. This reduces bias \textit{with} the error signal, and suppresses its effects.}
\end{figure}

\begin{figure}
\includegraphics[width=8.4cm]{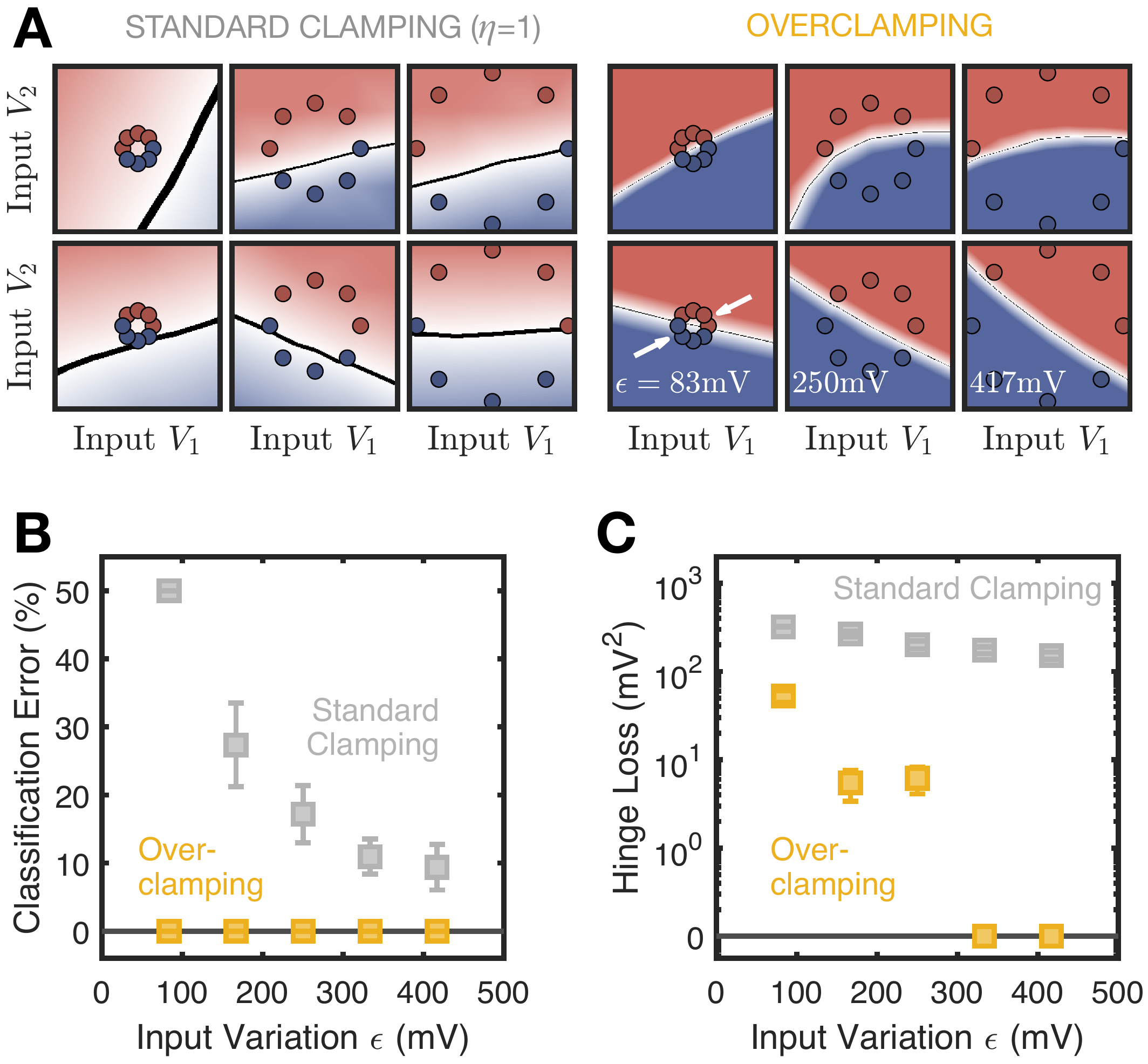}
\caption{\label{fig:fig5} \textbf{Overclamping with Full Experimental Network} \textbf{A} Classification results for the two methods. Background color is output after training, dots are training data. Both methods use cycle time $\tau = 200~\mu$s, further details in Materials and Methods. The data is divided differently in each row, and columns are in order of increasing input variation $\epsilon$.
\textbf{B} Classification accuracy for each method as a function of input variation $\epsilon$. Each data point is an average of the 8 experiments of the same task with rotated labels. Error bars are standard error.
\textbf{C} Hinge loss as a function of input variation. Averaging same as in B. Log axis breaks to include 0 (black line). }
\end{figure}

Our modified protocol, \textit{overclamping}, reduces the impact of bias simply by modifying the amplitude and duration ($t_h$) of the clamped boundary condition for the training set. 
In Coupled Learning~(Eq.~\eqref{expRule}), the output voltage is nudged proportionally to the signed error $\delta_i=L_i-O_i^F$. As the system approaches the solution, $\delta_i \rightarrow 0$ so the bias term dominates. This has the effect of shifting the learning landscape (resulting in an error floor) and pushing the system along basins of zero error, which creates the demonstrated cyclic dynamics. To avoid these problems, we nudge in the same direction but by a constant value $\eta L_0$:
\begin{equation}
    O^C_i=O^F_i+ \eta L_0 \text{ sign}(\delta_i).
\end{equation}
For a good choice of $\eta L_0$, the voltage difference in the learning rule $ (V_F^2 - V_C^2 )$ is large enough to dominate the bias $B$.
To ensure convergence and stability near zero error, we implement an error-dependent step time $t_h = t_0 |\delta|$. This technique is sketched visually alongside standard clamping in Fig.~\ref{fig:fig4}. We modify $t_0$ during training to keep the period ($\tau$) approximately constant. In theory, these modifications give qualitatively identical results to standard Coupled Learning, but in practice are far less sensitive to the presence of a nonzero bias, see Materials and Methods for derivation and details.

We demonstrate results of the overclamping protocol for a set of binary classification tasks.
We utilize the same method for classification as in Fig.~\ref{fig:fig1}, wherein outputs exceeding their labels are considered to have zero error.
For overclamping this amounts to $t_h=0$ and thus simply skipping these data in training. 
With this different relationship to training time, a direct comparison of the steady-state results of Fig.~1 is difficult, however analysis of the final results is directly comparable.
We note also that overclamping shines with smaller labels and small input variation, whereas standard clamping requires both to be larger to even sense the error signal. To highlight a general comparison,
we train the system to classify eight datapoints arranged in a circle (with trials for five different radii) into two classes, as shown for several examples in Fig.~\ref{fig:fig5}A.
Coupled Learning (standard clamping) learns passably when input values are widely separated, but as the diameter $\epsilon$ of the training data in input space shrinks, performance rapidly degrades, as shown in Fig.~\ref{fig:fig5}B and C. In contrast, overclamping not only succeeds (0\% classification error), but creates a far sharper boundary (lower hinge loss), possibly because the signal does not vanish at the solution. 

\section{Discussion}
Analog signals have the potential to be faster and more energy-efficient than their digital counterparts, but are sensitive to physical imperfections. In this work we have studied this sensitivity in contrastive local learning networks (CLLNs), approaching it as a fundamental characteristic rather than an engineering flaw. We showed that in our system, imperfection in physical components generates an uncontrolled bias in learning dynamics, leading to limit cycles in the dynamics instead of the fixed points expected for perfect systems. We developed an analytical model that quantitatively matches the experimental dynamics and suggests that such behaviors will emerge from bias of \textit{any} magnitude, making them the rule in analog systems, not the exception. 

Our results suggest intriguing comparisons between overparameterized analog systems and underparameterized ``perfect" (\textit{e.g.} simulated) ones. The cyclical dynamics we observe even when a joint solution is available (Fig.~\ref{fig:fig2}) are reminiscent of those seen when training for two incompatible tasks (under-parameterized) in a variety of bias-free systems~\cite{falk_learning_2023}. In our linear theory, bias `tilts' the landscape, removing regions of joint stability between compatible tasks. This is why the system behaves as if they are not mutually satisfiable. Further work is required to determine if this intuition extends to complex nonlinear landscapes, state-dependent bias, noise, and other types of tasks like classification.

Our work opens several new venues of inquiry. First, we note that simulated (and hence ideal) physical learning machines spectrally entangle the physical and learning error landscapes~\cite{stern_physical_2024}, pairing soft physical modes with stiff error modes ~\cite{stern2025physical,guzman2025microscopic}. It is unclear how these relationships are affected by noise and bias. Do the limit cycles observed here suppress, preserve, or enhance the physical signatures of the learned solutions? 

Second, the effects of noise require more study; in particular, it is important to understand in what contexts noise may be beneficial. Noise on parameter updates has been shown to increase robustness in deep feed-forward digital~\cite{neelakantan2017adding} and analog networks~\cite{ye2023improving}. Similar effects are seen when randomizing inputs (stochastic gradient descent)~\cite{goldt_dynamics_2020,feng_inverse_2021} or randomly dropping connections (dropout)~\cite{srivastava_dropout_2014}. Noise in edge update timing has been shown to improve solutions in linear self-learning analog networks~\cite{wycoff_desynchronous_2022}, but the impact of update noise in these systems is unknown. Currently, noise is negligible for our macroscopic electronic components but could become crucial in networks with more components that are microfabricated, or that use different resistive elements like memristors. In such systems, noise could arise both in the updates and node voltages. The effects of and interplay between these sources of noise in Contrastive Local Learning Networks is an important avenue for future work. Importantly, overclamping also can be used to suppress the effects of noise by amplifying the learning signal.

Third, we have demonstrated some effects of bias but others remain to be explored. In particular, while we have shown that error increases with bias, it is possible that some bias may be beneficial for generalization. A uniform bias towards zero (an $L_1$ regularizer) can help prevent overfitting in digital networks~\cite{mehta_highbias_2019}, while an uncontrolled bias is often harmful~\cite{ghosh_how_2023_2}. In our case, bias introduces further randomness in the projection of the Coupled Learning rule onto the direction of gradient descent. By adjusting the degree of overclamping to control the effects of bias, we could explore whether a small amount of this source of randomness allows the system to overcome barriers in the cost landscape to find more generalizable solutions.

We have demonstrated for classification tasks that the harmful effects of too much bias can be mitigated using our system-agnostic technique, \textit{overclamping}, without any modification to the physical components themselves.  
Our conceptual approach of controlling step size by step time rather than an ever-shrinking error signal can be useful for training a broad range of physical systems, where component variability, thermal fluctuations (noise), and other imperfections can inhibit training, including spintronics~\cite{kaiser_hardwareaware_2022}, memristors~\cite{ambrogio2014statistical, dalgaty2021situ}, and even mechanical systems~\cite{altman_experimental_2024}. Notably, in analog systems, the steps of measurement, calculation \textit{and} application of updates are all potentially noise- and bias-laden, and each benefit from enhanced signals. 
Furthermore, it is likely more energy-efficient than standard Coupled Learning, since the change in power from the modified output boundary conditions is small and varied in sign, but the time each state is run is nearly always reduced.
Moreover, overclamping is a broadly applicable technique that extends beyond Coupled Learning. By modifying the duration and magnitude of the boundary conditions, it can help counterbalance imperfect learning signals that hinder (physical) learning rules, such as directed aging~\cite{pashine_directed_2019}, equilibrium propagation~\cite{scellier_energybased_2023a} and its variants, and any training protocol that uses perturbations to extract gradient information, including of weights, as in direct feedback alignment or multiplexed gradient descent~\cite{mccaughan_multiplexed_2023}. 

We note that our approach differs from many standard efforts to reduce the impact of physical imperfection in learning and computation. One common approach is to build in redundancy in the hardware. This approach has been used, for instance, for fabricating extremely large AI chips~\cite{kundu_comparison_2025} or by spreading information across many physical entities as in quantum error correction~\cite{zhao_realization_2022}. A second is to identify, measure, and correct errors as they occur; in our system, this would amount to measuring bias in each scenario, then offsetting it. A third is simply to expend more effort creating extremely precise hardware components.
Unlike all these methods, our overclamping protocol requires no extra elements, additional measurements, or pristine components.
Nor does it rely on precise convergence between digital and physical systems, which often necessitates data-driven modeling~\cite{wright_deep_2022} or hardware modification~\cite{ambrogio2018equivalent}. \textit{All of these techniques}---redundancy, achieving uniformly high component or modeling precision, and extensive measurement---\textit{attempt to shoehorn analog learning and computation into a longstanding paradigm that implicitly relies on digital precision.} As a result, they each grow increasingly untenable at scale. 

It remains to be seen whether overclamping bears any similarity to approaches used by the brain. The bias-driven changes along solution spaces in our work are reminiscent of \textit{representational drift} in neural dynamics~\cite{driscoll2022representational,bauer2024sensory}, where the underlying firing patterns (but not the overall effect) change over time. This connection and others, including power-vs-precision tradeoffs~\cite{stern_training_2024} and clock-free training~\cite{wycoff_desynchronous_2022}, suggests that physical learning systems may provide a simpler platform for interrogating neural phenomena, where physical (imperfect) components manage to learn at scale.

\section*{Acknowledgements}
This work was supported by the UPenn MRSEC via NSF grant MRSEC/DMR-2309043 (SD, DJD) and NSF-DMR-MT-2005749 (MG, AJL). SD and MG acknowledge support from both the Data Driven Discovery Initiative and the Center for Soft and Living Matter at the University of Pennsylvania.
\section*{Data Availability}
All data  and software are publicly available at \url{https://doi.org/10.5281/zenodo.15692914} \cite{repo}.

\appendix
\section{Error Definitions} \label{ErrorCalc}

We define error for a task $\alpha$ at time $t$, $E_\alpha(t)$, to be the average of output-label deviations squared across of all $N$ data points in the task:
\begin{equation}
    E_\alpha(t) = \frac{1}{N} \sum_{i=1}^N \delta _i(t)^2
    = \frac{1}{N} \sum_{i=1}^N (L_i-O_i(t))^2,
\end{equation}
where as in the main text, $\delta_i(t) = L_i-O_i(t)$.
For a classification task, we modify this equation to not penalize the system for `very correctly classified' data points. Specifically,
\begin{equation}  E^{\text{class}}_\alpha(t) = \frac{1}{N} \sum_{i=1}^N \delta_i(t)^2  \Theta \big (L_i\delta_i(t)\big)
\end{equation}
where $\Theta$ is the Heaviside step function, zeroing out any data point for which $L_i$ has the same sign as $-\delta_i$.

When an experiment involves two tasks ($\alpha$, $\beta$) we define the error as the average of the two task errors:
\begin{equation}
    E(t) = \frac{1}{2} \left (E_\alpha(t)+E_\beta(t)\right )
\end{equation}
When such an experiment cycles between two tasks with period $\tau$, we define the \textit{combined error} as the average of this error at times $t=0=\tau$ and $t=\tau/2$ after the system has reached its limit cycle:
\begin{equation}  
    \overline E = \frac{1}{2} \big (E(\tau/2)+(E(\tau)\big ) \Big |_ {\text{limit cycle}}
\end{equation}

When training for binary classification, (\textit{e.g.} Fig.~\ref{fig:fig1}C), each class ($+$,$-$) has a single label with $L_+ = -L_-$. 
The training protocol is applied as normal, except when an output value exceeds its label in the \textit{desired} direction, $\left ( O_F - L \right ) \times\text{sign}(L) > 0$. In this scenario $\eta$ is set to 0, making $O^C=O^F$ and (in theory) driving no change in the network. This protocol effectively creates a shifted \textit{hinge loss}, useful for classification problems.

\section{Task Data} \label{taskdata}

In all figures, task $\alpha$ is blue and $\beta$ is red. $V_+=0.435$~V, $V_-=0.018$~V. \\

\noindent
\textbf{Fig~\ref{fig:fig1}C-D: }
Labels for task $\alpha$ are all $L_+ = 63$~mV, and task $\beta$ are $L_- = -63$~mV. Inputs are arranged in a square grid in 2D input space [$V_1$,$V_2$] as shown.

\noindent
\textbf{Fig~\ref{fig:fig1}E-F: }
[$V_1$, $L$]:  $\alpha$ [$V_+$, 0.021]V and $\beta$ [$V_-$, -0.063]V. \\

\noindent
\textbf{Fig~\ref{fig:fig2}B,F: }
[$V_1$, $L$]:  $\alpha$ [$V_+$, 0.31]V and $\beta$ [$V_-$, 0.14]V.

\noindent
\textbf{Fig~\ref{fig:fig2}C,G: }
[$V_1$, $L$]:  $\alpha$ [$V_+$, 0.31]V and $\beta$ [$V_-$, 0.18]V.\\

\noindent
\textbf{Fig~\ref{fig:fig3}: }
[$V_1$, $L$]:  $\alpha$ [$V_-$, 0.143]V and [$V_+$, 0.268]V and $\beta$ [$V_-$, 0.156]V.\\

\noindent
\textbf{Fig~\ref{fig:fig5}: }
Labels for task $\alpha$ are all $L_+ = 18$~mV, and task $\beta$ are $L_- = -18$~mV. Overclamping uses $L_0=V_+$, $\eta = 0.25$. Standard clamping uses $\eta = 1$. Inputs are arranged equally spaced around a ring, centered in 2D input space [$V_1$,$V_2$] as shown, with radii equally spaced between $\epsilon=$83~mV to 417~mV. Data in panels B and C denote the mean and standard error across eight trials, one with every possible linear split of these data locations (corresponding to rotating the dividing line between red and blue points). Two of the eight such divisions are shown in panel A.

\section{Learning Rule and Bias}\label{biassec}

The constant $A_0$ in the learning rule (Eq. \ref{expRule}) is consistent across edges and results from our choice of electronic components. `Perfect' functioning of the integrated circuits enacting the learning rule would result in
\begin{equation}
    A_0 \equiv \frac{1}{V_0 R_0 C_0} = \frac{1}{ (0.33~\textrm{V})(100~\text{ohms}) (22~\mu\textrm{F})} \approx \frac{1.4}{\textrm{V~ms}}
    \label{A0_exp2}
\end{equation}
for the experiments in Fig.~\ref{fig:fig2} and
\begin{equation}
    A_0 = \frac{1}{(0.33~\textrm{V} (1~\textrm{k~ohms})(2.2~\mu\textrm{F})} \approx \frac{1.4}{\textrm{V~ms}}
        \label{A0_exp3}
\end{equation}
for all other experiments. Note that $C_0$ is the capacitor holding the gate voltage $G$ and that $R_0$ sets the scale of the charging current. See \cite{dillavou_machine_2024} for details.

In practice, this rate deviates slightly from its predicted value. We measure $A_0$ by enforcing and measuring a variety of voltages on all four terminals of a pair of twin self-adjusting resistors, and measuring the resulting change in gate voltage. By comparing predicted (Eq.~\ref{expRule}) and actual change, we find that for the experiments of Fig.~\ref{fig:fig2}  (Eq.~\ref{A0_exp2}), $A_{\exp} \approx 2 A_0$, and for the remaining experiments (eq. \ref{A0_exp3}), $A_{\exp} \approx A_0 \text{ to } 1.5 A_0$. The deviation likely comes from using $R_0$ out or nearly out of spec for the analog multiplier, as $R_{\min} = 1~\textrm{k~ohms}$. We use the appropriate value of $A_{\exp}$ to generate our model predictions in Figs~\ref{fig:fig2} and \ref{fig:fig3}. We note that the only effect of this change is to accelerate the dynamics.

The bias $B$ in Eq. \ref{expRule}, however, does alter the dynamics, as we show in the main text. It originates from imperfect aspects of the electronic components, and likely as a result of the effects detailed below. For each edge the bias is an unknown, likely complicated but deterministic function of the system state. We note that each source of bias listed below scales as $(R_0 C_0)^{-1}$, making a change in these components unhelpful in reducing bias. For simplicity we choose $R_0=1~\textrm{k~ohms}$, $C_0=2.2~\mu\textrm{F}$ for calculations below, and note that doing so has no effect on the relative size of the bias sources to each other or the learning rate.

\textbf{Non-identical Twin Networks:} One potential source of bias is the discrepancy between free and clamped networks. For a variety of voltages applied at inputs, in the absence of clamping at the outputs, the free and clamped networks should have the same set of node voltages. We find deviations in node voltages of order 1~mV, with typical voltage drops of order 100~mV. To probe the maximum effect of this bias we assume the system is in a state where the learning rule \textit{would be} zero in a perfectly mirrored system (signal = 0). However, discrepancies create deviations such that $V_F \approx V_C + 1$~mV. Plugging this into \eqref{expRule} we have:
\begin{equation}
    \Delta G = t_h \times A_0 \Big (201~(\textrm{mV})^2\Big) = t_h \times (275~\textrm{mV/s}),
\end{equation}
which gives us a scale of drift for $B$ that is about an order of magnitude smaller than what we observe experimentally. Discrepancies between twin edges are therefore not the leading cause of bias.

\textbf{Imperfect Voltage Drop Measurements:} A second potential source of bias is imperfect voltage drop measurements across each edge, which comes down to component error of operational amplifiers and a multiplier circuit. These imperfections would generate error in the same manner as the previous calculation, and are of the same order (1\% error) based on the specifications of the integrated circuits used (TLV274IN and AD633ANZ).

\textbf{Imperfect Execution of Learning Rule:} The third and likely dominant source of bias is the output offset voltage of the multiplier chip (AD633ANZ), which generates the charging current. This bias voltage (listed as $V_B\approx$ 5~mV) directly creates a current $V_B/R_0\approx 5~\mu\textrm{A}$ onto the capacitor. Then we have:
\begin{equation}
    \Delta G = t_h \times \frac{V_B}{R_0 C_0} = t_h \times (2.3~\textrm{V/s})
\end{equation}
which is the order of the drift we see in experiments.

An additional source of bias in executing the learning rule is the current draw of the summing port of the multiplier, which is of order $1~\mu\textrm{A}$, making it comparable to the other sources of bias, but definitively overshadowed by the output offset voltage.

\section{Theory}\label{theorysec}
The theoretical model relies on three approximations, detailed below and explained for a single task.
The first is a continuum description of the evolution of gate voltages,
\begin{equation}
    \frac{d\mathbf G}{dt}=-\gamma\frac{\partial \mathcal C}{\partial \mathbf G}+\mathbf B,
\end{equation}
where $\gamma=A_0/S$ is the learning rate. As described in the previous section, $A_0$ deviates from its predicted value depending on a choice of electronic components. As a result we measure this effect and use the resulting $\gamma \approx 2.4\times 10^6~\text{ohm}/\textrm{s}$ for the theory results of Fig.~\ref{fig:fig2}, and $\gamma=3.2\times 10^6~\text{ohm}/\textrm{s}$ for Fig.~\ref{fig:fig3}.

Second, we linearize the learning dynamics around a minimum $\mathbf G^*$ of the contrastive function, i.e. a learning solution,
\begin{equation}
    \frac{d\mathbf G}{dt}=-\mathbf M(\mathbf G-\mathbf G^*)+\mathbf B.
    \label{eq:lineardynamics}
\end{equation}
Here, $\mathbf M =\gamma \frac{\partial^2 \mathcal C}{\partial \mathbf G\partial \mathbf G}\Big|_{\mathbf G^*}$, is proportional to the Hessian of the contrastive function evaluated at the solution.
The choice of $\mathbf G^*$ is determined by the next approximation.

Third, we linearize the physical dynamics around the zero voltage state---for this calculation we emphasize that we consider linear resistor networks.
We recall that the conductance of a single edge $i$ (Eq.~\eqref{conductance}) connecting nodes $a$ and $b$ depends on the voltage at both terminals, $K_i=S(G_i-V_T-0.5(V_a+V_b))$, where $S$ and $V_T$ are fixed values.
The power dissipated by this edge, expanded to quadratic order in voltages, corresponds to $K_i(V_a-V_b)^2\approx S(G_i-V_T)(V_a-V_b)^2$.
In other words, in the linear regime conductances become independent of the voltages $V_a$ and $V_b$.
Moreover, the voltage response to some fixed voltage inputs is defined by Kirchhoff's  current law, and corresponds to the minimization of the power dissipation $\mathcal P=\sum_i S(G_i-V_T)(\Delta V)_i^2$.
This is approximately correct and equivalent to the analytical treatment given to the same system in~\cite{dillavou_machine_2024}.
Following \cite{guzman2025microscopic}, the linearized physical response to external current  is simply
\begin{equation}\mathbf V_F=\mathbf H^{-1}\mathbf I,
\label{eq:linresponse}
\end{equation}
where $\mathbf V_F=(V_1,V_2,V_3,...)$ is the vector of node voltages, $\mathbf I=(I_1,I_2,..)$ is the vector of external current applied to the nodes and needed to enforce a voltage boundary condition, and $\mathbf H$ is the Hessian of the power ($\partial^2 \mathcal P/\partial V_i\partial V_j$) when having at least one fixed voltage node, ensuring that $\mathbf H$ is invertible.

As a result, the contrastive function becomes an explicit yet lengthy function of the gate voltages,
\begin{equation}
    \mathcal C =\mathcal P_C-\mathcal P_F = \mathcal P[\mathbf V_C(\mathbf G)]-\mathcal P[\mathbf V_F(\mathbf G)],
\end{equation}
from which we can directly compute the contrastive Hessian $\mathbf M$.
For more details see~\cite{guzman2025microscopic} and the code repository~\cite{repo}.

In practice, we choose $\mathbf G^*$ to be the point in the solution space of the linear resistor network that lies closest to the location in gate voltage space where the bias is measured (see next section).

The linear evolution in Eq.~\eqref{eq:lineardynamics} has an analytical solution given by
\begin{align}
    \mathbf G(t,\mathbf G_0) = e^{-\mathbf M t}\mathbf G_0+\left[\int_0^t e^{-\mathbf M s}ds\right]
(\mathbf M \mathbf G^*+\mathbf B),
\end{align}
where $\mathbf G_0$ is the initial gate voltage state and $e^{-\mathbf Mt}$ denotes the matrix exponential of $-\mathbf Mt$.
Considering that $\mathbf M$ is a positive semi-definite matrix, having only positive or zero eigenvalues, we can further decompose the previous relation in terms of its (pseudo)inverse $\mathbf M^\dagger$ and projector onto its nullspace $\mathbf N$
\begin{equation} 
    \mathbf G(t,\mathbf G_0)=e^{-\mathbf M t}\mathbf G_0+(\mathbf 1-e^{-\mathbf M t})(\mathbf G^*+\mathbf M^\dagger \mathbf B)+t\mathbf N \mathbf B.
    \label{eq:linearsolution}
\end{equation}
Notice that first two terms correspond to exponential decays, reaching a fixed value in the limit of $t\rightarrow \infty$, whereas the last term represents a constant drift induced by the bias along its projection onto the nullspace. \\

\begin{figure}
\includegraphics[width=1\linewidth]{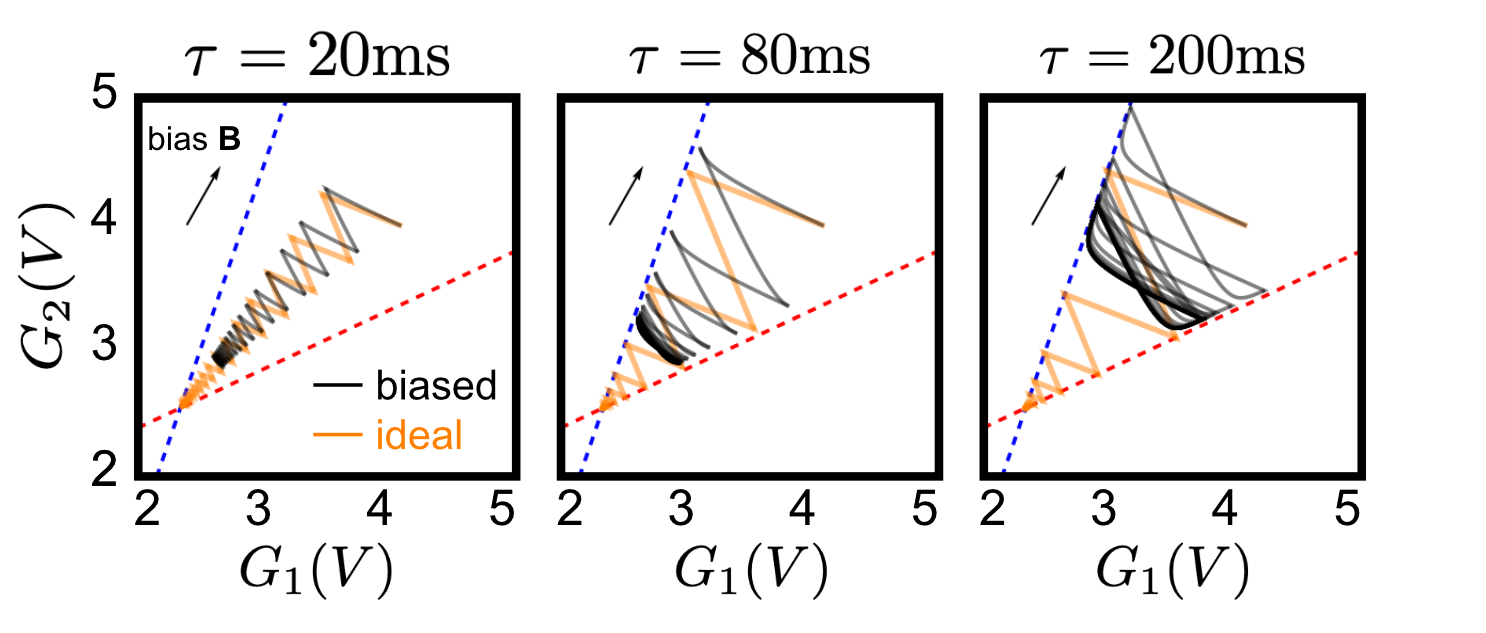}
    \caption{{\bf Transient dynamics of biased and ideal systems.} Learning trajectories in the linear approximation for a biased (black) and ideal (yellow) systems, for three different training periods: $\tau =20\text{ms}$ (left), $80\text{ms}$ (middle), and $200\text{ms}$ (right). The parameters of the network, together with the measured biased (black arrow), are the same as those of Fig.~\ref{fig:fig2}F. For each case, we use the same initial condition $(G_1,G_2)=(4\text{V},4\text{V})$ and computed the evolution for 20 learning cycles, i.e. a  total time of $T=20\tau$.}
    \label{fig:transient}
\end{figure}

\textbf{Periodic training:}
Equation~\eqref{eq:linearsolution} remains valid for periodic training that alternates between tasks $\alpha$ and $\beta$.  At each half-period both the Hessian and the solution change with the task, $M_{\mu}$ and $\mathbf G^*_\mu$, with $\mu =\{\alpha,\beta\}$.
Denoting the evolution of the gate voltages under each task by $\mathbf G_\mu(t)$, the analytical solution under periodic training is found by imposing continuity in time,
\begin{equation}
    \mathbf G_{\alpha}(\tau/2,\mathbf G_0) = \mathbf G_\beta(0,\mathbf G_\alpha(\tau/2,\mathbf G_0)),
\end{equation}
i.e. the gate voltages at the end of task $\alpha$ are the same as those at the beginning of task $\beta$.\\
In Fig.~\ref{fig:transient} we compare the transient dynamics of the biased and ideal systems modeling the experimental setup of Fig.~\ref{fig:fig2}F for three values of $\tau$. Starting from the same initial condition, the learning trajectories in the absence of bias (ideal) reach the joint solution by zigzagging between the solution lines of each task. In contrast, the biased trajectories cannot approach the joint solution and instead get trapped in limit cycles.

\textbf{Non-equilibrium steady cycles:}
Similarly, we can find the non-equilibrium steady trajectories shown in Figs.~\ref{fig:fig2} and~\ref{fig:fig3} by imposing periodicity.
We use $\bar{\mathbf G}_{\text{0}}$ to denote the state at the beginning of task $\alpha$ (once the system has reached the limit cycle).
By definition after a full period $\tau$ it reaches the same state:
\begin{equation}
    \mathbf G_{\beta}\left(\frac{\tau}{2},\mathbf G_{\alpha}\left(\frac{\tau}{2},\bar{\mathbf G}_{\text{0}}\right)\right)=\bar{\mathbf G}_{\text{0}}.
    \label{eq:ness}
\end{equation}
We define the following variables for convenience
\begin{align}
    \mathbf A_{\mu} &\equiv e^{-\frac{\tau}{2}\mathbf M_{\mu}},\\
    \mathbf b_{\mu} &\equiv \left(\mathbf 1-e^{-\frac{\tau}{2}\mathbf M_{\mu}}\right)\left(\mathbf G^*_{\mu }+\mathbf M^\dagger_{\mu}\mathbf B\right)+\frac{\tau}{2}\mathbf N_{\mu}\mathbf B.
\end{align}
After which, Eq.~\eqref{eq:ness} becomes
\begin{equation}
    \mathbf A_\beta\left(\mathbf A_\alpha \bar{\mathbf G}_{\text{0}}+\mathbf b_\alpha\right)+\mathbf b_\beta=\bar{\mathbf G}_{\text{0}},
\end{equation}
with solution
\begin{equation}
    \bar{\mathbf G}_{\text{0}}=\left(\mathbf 1-\mathbf A_\beta\mathbf A_\alpha\right)^{-1}\left(\mathbf A_\beta \mathbf b_\alpha+\mathbf b_\beta\right).
\end{equation}
The inverse exists whenever there are closed orbits (e.g. Figs.~\ref{fig:fig2} and ~\ref{fig:fig3}), otherwise it should be taken as a pseudoinverse.
Finally, the evolution of the gate voltages over a full period can be found using $\bar{\mathbf G}_0$ as the initial condition into the dynamical equation~\ref{eq:linearsolution} for task $\alpha$, chained with those for task $\beta$.

\subsubsection*{Analytical solution for the small network}
For the three edge network analyzed in the main text (Fig.~\ref{fig:fig2}A), we can explicitly write down the inverse of the Physical Hessian in terms of the gate voltages and solve for the output voltage according to Eq.~\eqref{eq:linresponse}. Alternatively, for such a small system we can directly solve by imposing current conservation.
Using the notation of the main text (see Fig.~\ref{fig:fig2}A), we have $\mathbf V_F=(V_+,V_-,V_1,O)$, where the first two entries are the high and low ground voltages, the third is the input voltage, and $O$ is the voltage at the output node given by
\begin{equation}
    O=\frac{V_1 R_0^{-1}+S(G_2V_++G_1 V_--(V_-+V_+)V_T)}{R_0^{-1}+S(G_2+G_1-2V_T)},
\end{equation}
where $G_1$ is the gate voltage of the transistor connecting $O$ to $V_-$, $G_2$ from $O$ to $V_+$, and $R_0$ is the fixed resistance of the edge connecting $V_1$ to $O$.

\section{Measuring Experimental Bias}\label{CalculatingExpParams}

For each experiment, the bias vector $\mathbf B$ is inferred from the gate voltage dynamics of individual tasks.
For each single-point task of Fig.~\ref{fig:fig2}, we run independent experiments, one per task, to measure the bias.
We look at the long time dynamics in non-periodic setups, governed by the drift along the solution line as shown in the main text.
The rate of change of the gate voltages at long times then corresponds to projection of $\mathbf B$ along the solution line, $d\mathbf G/dt\approx \mathbf N\mathbf B$ from Eq.~\eqref{eq:linearsolution}.
Experimentally, the rate is not constant and to compare with the linear theory we must choose where to measure this rate. 
For each task, we choose a point along the solution line that is closest to a predetermined central point to the observed dynamics.
These central points are [$G_1$,$G_2$] = [3.0, 3.5] V in Fig.~\ref{fig:fig2} F, [3.1, 4.3] V in Fig.~\ref{fig:fig2} G, and [2.5, 2.5] V in Fig.~\ref{fig:fig3}, see ~\cite{repo}.

For the two tasks in Fig.~\ref{fig:fig3}, we inferred the bias from the periodic trajectories.
We used the same method mentioned before for the single-point task $\beta$.
For task $\beta$, consisting in two data points, there is no direction of zero curvature and the bias can be completely determined by setting   $d\mathbf G/dt=0$ in Eq.~\eqref{eq:lineardynamics}: $\mathbf B = \mathbf M(\mathbf G-\mathbf G^*)$, where $\mathbf G$ is the stationary point obtained from the experiments. However, such measurement relies on the imprecise values of all the curvatures through the matrix $\mathbf M$.
Instead, we consider only the projection along the smallest curvature direction (the center line depicted by the blue trajectory in the last panel of  Fig.~\ref{fig:fig3}C).

Then, for each experiment, we find the unique value of $\mathbf{B}$ that satisfies the projections found for each of the two tasks. These values are $\mathbf {B} = (2.9, 4.7)$ V/s in Fig.~\ref{fig:fig2}F, $(-0.3, 2.0)$ V/s in Fig.~\ref{fig:fig2}G, and $(-1.8, 0.6)$ V/s in Fig.~\ref{fig:fig3}. 

We note one change between here and previous work, which is the removal of a drainage resistor (34~\textrm{k~ohms} in \cite{dillavou_machine_2024}~Fig 7), designed to funnel the capacitor charging current to ground when learning is turned off. This resistor biased learning updates, effectively shuttling charge between the capacitor and ground when learning is toggled. We reintroduce this resistor only in Fig.~\ref{fig:fig2}
C and G, and only for a single network edge ($G_1$), in order to explore the effect of varying bias.

\section{Clamping Protocols}\label{overclampingsec}
\textbf{Standard Clamping:}
In standard clamping, the clamped state corresponds to a slight perturbation to the free state, in which the output voltage values are nudged in the direction of the labels (Eq.~\eqref{clamping}).
In terms of the signed error $\delta=L-O^F$, the clamped output voltages are
\begin{equation}
    O^C=O^F+\eta \delta,
    \label{eq:nudge}
\end{equation}
leading to a clamped state 
\begin{equation}
    \mathbf V_C\approx \mathbf V_F+\chi\eta \delta+\mathcal O(\delta^2),
\end{equation}
where $\chi = \frac{d\mathbf V_C}{dO}(O^F)$.
The learning rule then becomes
\begin{align}
    \Delta \mathbf G &= t_h\left [ \mathbf B +A_0 \Big (\mathbf V_F^2 - \mathbf V_C^2\Big ) \right ] \\
    &=  t_h\left [ \mathbf B -2A_0\mathbf V_F\chi\eta \delta \right]+\mathcal O(\delta ^2),
    \label{eq:standardclamping}
\end{align}
which shows explicitly that the learning dynamics are dominated by bias as the error approaches zero. \\

\textbf{Overclamping:}
To overcome the bias we implement two modifications.
The first one is to make the learning signal stronger to compensate the bias.
We do so by 
adapting the nudging protocol in Eq.~\eqref{eq:nudge}.
Instead of nudging towards the desired value $L$, we nudge towards a value constant magnitude, $L_0\text{ sign}(\delta)$, with $L_0\gg|L|,|O^F|$, \textit{i.e.} to a label in the right direction but far away.
As a result,
\begin{align}
    O^C&=O^F+\eta (L_0\text{ sign}(\delta)-O^F) \\&\approx O^F  +\eta L_0 \text{ sign}(\delta),
\end{align}
and the clamped state has a zeroth order contribution,
\begin{equation}
    \mathbf V_C\approx \mathbf V_F+ \mathbf C \eta\text{ sign}(\delta)+\mathcal O(\delta),
\end{equation}
with $\mathbf C$ a constant vector proportional to $L_0$.
However, as the error goes to zero, the clamped state does not approach the free state.
The second change, to ensure convergence, is changing $t_h$, the time the system applies the learning rule per sample, proportional to the error, $t_h=t_0|\delta|$, and $t_0$
The overclamping learning rule, in the small error limit, then becomes
\begin{equation}
    \Delta\mathbf G = t_0 |\delta|\left(\mathbf B-2 A_0\mathbf V_F \mathbf C\eta\text{ sign}(\delta)\right)+\mathcal O(\delta^2).
\end{equation}

We choose $\eta=0.25$, $L_0 \approx 0.43$~V, $|L_{\pm}|= 18$~mV for Fig.~\ref{fig:fig5}.

\bibliography{mergedbib}

\end{document}